\documentclass[11pt]{JHEP3}
\usepackage{amsmath}
\usepackage{amsfonts}
\usepackage{amssymb}
\usepackage{cite}
\allowdisplaybreaks

\title{The algebra of flat currents for the string on $AdS_{5}\times
S^{5}$ in the light-cone gauge}

\author{Ashok Das, A. Melikyan and Matsuo Sato\\
  Department of Physics and Astronomy, University of Rochester\\
  Rochester, NY 14627-0171, U.S.A.\\
  E-mail: \email{das@pas.rochester.edu},
          \email{arsen@pas.rochester.edu},
          \email{sato@pas.rochester.edu}}

\abstract{
We continue the program initiated in
\href{http://arxiv.org/abs/hep-th/0411200}{hep-th/0411200}
and calculate the algebra of the flat currents for the string
on $AdS_{5}\times S^{5}$ background in the light-cone gauge with $\kappa$ symmetry fixed. We find that the
algebra has a closed form and that the non-ultralocal terms come with a
weight factor $e^{\phi(\sigma)}$ that depends on the radial $AdS_5$
coordinate. Based on results in two-dimensional sigma models coupled
to gravity via the dilaton field, this suggests that the algebra of
transition matrices in the present case is likely to be unambigous.}


\preprint{}

\begin{document}

\section{Introduction}
\label{sec:intro}

Recently there has been much interest in studying the integrable
structures of the $\mathcal{N}=4$ SYM\ theory as well as of strings
propagating  on the $AdS_{5}\times S^{5}$
background with the goal of gaining a deeper insight into the $AdS/CFT$
correspondence~\cite{Maldacena:1997re,Gubser:1998bc, Witten:1998qj, Aharony:1999ti}.
In this context, significant progress has been made on
the SYM side in determining the anomalous dimensions of conformal
operators and comparing them with the energy spectrum of the string
states~\cite{Berenstein:2002jq, Minahan:2002ve, Bena:2003wd, Hatsuda:2004it,
Dolan:2003uh, Beisert:2003yb, Beisert:2003tq, Beisert:2003ea,
Arutyunov:2003rg, Arutyunov:2003uj, Arutyunov:2004xy, Beisert:2004yq,
Frolov:2003qc, Arutyunov:2004yx, Swanson:2004qa, Alday:2005jm,
Arutyunov:2005nk, Agarwal:2005ed, Agarwal:2005jj, Agarwal:2004cb,
Maharana:2005ny, Belitsky:2004yg, Belitsky:2004sc, Belitsky:2004sf} (for recent reviews see~\cite{Swanson:2005wz,
  Plefka:2005bk,
Beisert:2004ry, Tseytlin:2004xa, Tseytlin:2003ii, Belitsky:2004cz} and references therein).
It is becoming increasingly  evident from the recent developments
in this direction that the Bethe ansatz provides
the main scheme suitable for understanding the quantum integrability of this
system. On the gauge theory side, the Bethe ansatz is known up to
three-loop order in the $su(2)$ sector
and more recently has been generalized to all-loop order
in~\cite{Beisert:2004hm, Beisert:2005fw}.  On the string side,  it was clarified
in~\cite{Kazakov:2004qf, Beisert:2005bm} how classical Bethe type
equations appear in the $su(2)$ sub-sector of the string. Their relation
with that of the spin chain on the SYM side at one loop order is then
established in the scaling limit when the effective coupling constant $
\lambda /L^{2}\longrightarrow 0$. Intuitively, it is reasonable to expect that
in parallel to what happens in the gauge theory side, there exists some
discretized version of the Bethe ansatz for the quantum string
leading to correct results in the classical limit. Indeed, a major
progress in this direction
has been made in~\cite{Arutyunov:2004vx} where such a remarkable consistent
discretization has been constructed. The origin of this construction,
however, is not well understood, and it is not obvious how to derive such
discrete Bethe equations in general, although there exist some helpful
clues in the $su(2)$ sector considered in~\cite{Kazakov:2004qf}. Indeed, the
sigma-model string action on $AdS_{1}$ $\times S^{3}$ takes a particularly
simple form, and when subjected to the Virasoro constraints, this
model  coincides with the interacting two-spin model of
Faddeev-Reshetikhin which is known to be quantum
integrable~\cite{Faddeev:1985qu}.  In
the $su(2)$ sub-sector of the string, it is obvious that the Virasoro
constraints play a crucial role. It is not clear, however, whether a
similar simplification may arise for the string on $AdS_{5}\times
S^{5}$ as well.

It is well known that sigma models (both bosonic as well as
supersymmetric) lead to a classical current algebra (Poisson bracket
algebra) that involves Schwinger terms (derivatives of delta function
which are also known as non-ultralocal terms).
In an earlier paper~\cite{Das:2004hy} the presence of
 such terms in the classical current algebra has
been explicitly verified for the string on $AdS_{5}\times S^{5}$. Even
if one sets the fermions of the theory to
zero (thus temporarily avoiding the technical difficulties associated
with the local  $\kappa $
-symmetry), the classical current algebra which has a closed form
contains  Schwinger
(non-ultralocal) terms. It is known~\cite{deVega:1983gy} that in the
principal chiral model in flat space-time, such terms lead to
difficulties (nonuniqueness) in the calculation
of the algebra of the monodromy matrices which is essential for the
quantization of the model and generally one needs a regularization
procedure to define this algebra. One interesting (but not
well-understood) method to deal with such difficulties was proposed by
Faddeev and Reshetikhin ($FR$)~\cite{Faddeev:1985qu} and involves
reducing (by hand) the
original theory to the one of the two interacting spins where the
classical current algebra does not have non-ultralocal terms. The
original classical current algebra is then argued to be recovered in a
rather special limiting procedure in the quantum version of the
reduced model. In the $su(2)$ sub-sector considered in~\cite{Kazakov:2004qf},
the Virasoro constraints effectively remove these non-ultralocal
terms. One may, then, ask whether the
Virasoro constraints alone are powerful enough to provide a consistent
reduction of the  full  $AdS_{5}\times S^{5}$ sector to a $FR$ type
model, and whether this depends in any way on the gauge-fixing scheme.
The answer to this is not at all obvious because of subtleties in gauge
fixing in a curved background. We would also like to point out here
another relevant interesting phenomenon which
occurs in the 2-dimensional sigma model coupled to gravity through the
dilaton field~\cite{Korotkin:1997fi}. Here, due to the fact that the
spectral parameter is a
local function of the world-sheet and explicitly depends on the dilaton field
through $\rho (x)\sim e^{\phi }$, the non-ultralocal terms in the
current algebra are manageable and the
algebra of the monodromy matrices is well defined. Basically, in this
case the dilaton field regularizes the difficulties arising from the
non-ultralocal terms in the classical current algebra. However,
because of conformal invariance, the string on $AdS_{5}\times S^{5}$
has no coupling to the dilaton and as a result this simple feature of
sigma models on a curved background cannot be directly used in the
calculation of the algebra of the monodromy matrices.
In this note we will explore the integrability properties
of the string on $AdS_{5}\times S^{5}$ background in the light-cone
gauge and attempt to explicitly calculate the current algebra. We will
show that in the light-cone gauge similar effective dependence on the
radial coordinate of $AdS_5$ also arises in the algebra of currents for the string on
$AdS_{5}\times S^{5}$ which may be helpful in the calculation of the
algebra of the monodromy matrices (and, therefore, in the quantization
of the model).

One of the main motivations for choosing the light-cone gauge is to
simplify the covariant superstring action thereby making it more
amenable  for algebraic calculations.  The covariant
Green-Schwarz formulation of the string on $AdS_{5}\times S^{5}$
background~\cite{Metsaev:1998it}, for example, is a non-linear sigma model on
the supercoset $\frac{PSU(2,2|4)}{SO(4,1)\times SO(5)}$ and the action
is rather formidable. For instance, because of $\kappa$ symmetry as
well as reparameterization invariance, the action contains fermionic
terms up to order $\theta^{32}$ which makes it rather impractical. It
is therefore necessary to fix these local symmetries. Indeed,
as shown in~\cite{Kallosh:1998ji, Kallosh:1998nx} the superstring on
$AdS_{5}\times S^{5}$
background significantly simplifies if one employs the Killing gauge for the
$\kappa $-symmetry: the resulting action will be limited to order
$\theta ^{4}$ in the fermionic terms. Past experience
with strings propagating in a flat background suggests that the
light-cone gauge may be the most appropriate if
one has to quantize the string on $AdS_{5}\times S^{5}$ background\footnote{Light cone gauge is also useful in the study of the $\mathcal{N}=4$ super Yang-Mills theory, which can be formulated on superspace in the light cone gauge. By using such a formulation, integrability of the anomalous dimension in the SL(2) sector at one and two loops has been studied for SYM theory in ~\cite{Belitsky:2004yg, Belitsky:2004sc, Belitsky:2004cz, Belitsky:2004sf}.}
This can be easily seen from the point of view of the Hamiltonian
analysis of the covariant superstring. The fermionic constraints of
the theory resulting from the $\kappa
$-symmetry lead to a complicated mixture of the first and
second class constraints which are impossible to disentangle
covariantly. As a result,  the fundamental Dirac brackets for the
theory are hard to construct\footnote{
There has been, however, some progress in covariant quantization using
the \emph{pure spinor formalism }(for a review
see~\cite{Berkovits:2002zk}.)}
On the
other hand, the fundamental Dirac brackets are absolutely necessary to
determine the classical $r$-matrix. The only solution, therefore, is
to fix appropriately the $\kappa $-symmetry and this can be conveniently
done in the light-cone gauge.

Our main goal lies in calculating the algebra of the flat currents
depending only on the physical fields (variables). It is clear from the above
discussion, that in order to do this, we must
fix both the bosonic reparameterization invariance (with the
light-cone gauge) as well as the local $\kappa $-symmetry. The
consistent light-cone formulation of the string on $AdS_{5}\times S^{5}$
background has been discussed by Metsaev and Tseytlin
in~\cite{Metsaev:2000yf} and further
elaborated using phase space Lagrangian in~\cite{Metsaev:2000yu}. We
mention here only one
important feature of this gauge fixing that will play a significant
role in our calculations. Namely, unlike
in the flat space case, it
is impossible to fix simultaneously the conventional bosonic light-cone
gauge $x^{+}=\tau $ and  $g^{\mu\nu}=\eta^{\mu\nu},
\mu,\nu=0,1$
consistent with the equations of motion for the string on the
$AdS_{5}\times S^{5}$ background~\cite{Horowitz:1990sr, Rudd:1994ss}. Instead
one can choose $x^{+}=\tau ,\mathcal{P}^{+}= {\rm const}$ in
the phase space, which translates into $h^{00}=-p^{+},$ $h^{11}=\left(
p^{+}\right) ^{-1}e^{4\phi },$ $h^{01}=h^{10}=0$ $\left( h^{\mu\nu } \equiv
\sqrt{-g}g^{\mu\nu }e^{2\phi }\right)$, in the coordinate space. As
discussed in~\cite{Alday:2005gi} the Virasoro constraints do
not follow from  the Lax representation of the sigma model and have to
be considered separately, and this particular gauge
choice corresponds to solving the Virasoro constraints explicitly. It is this
feature of the light-cone gauge that leads to the appearance of the $e^{\phi
}$ factor in the non-ultralocal terms in the current algebra.

The paper is organized as follows. In section~{\bf \ref{sec:review}} we
briefly summarize the main properties of the superstring on the $AdS_{5}\times S^{5}$ background.
In section~{\bf \ref{sec:lightcone}} we review the necessary results in the
light-cone gauge that are used in our calculations. In section~{\bf \ref{sec:algebraofcurrents}} we present
the algebra of currents without the spectral parameter in the
light-cone gauge and show that is has a closed form. In section
~{\bf \ref{sec:algebrawithspectral}} we present the algebra of currents with
the spectral parameter in the light-cone gauge. Here, we find that, in
general, the algebra is also closed. We also point out that potentially
dangerous non-ultralocal terms in the algebra come with a weight
factor proportional to $e^{\phi(\sigma)}$. This, therefore, is likely to lead to an unambiguous
algebra for the transition matrices. In Appendices {\bf A}-{\bf E}
we present all the technical details that are not given in the main text.

\section{Review of the Superstring on $AdS_{5}\times S^{5}$}
\label{sec:review}

We summarize here some of the basic properties of the type IIB Green-Schwarz
superstring action on the $AdS_{5}\times S^{5}$ background
~\cite{Kallosh:1998ji, Kallosh:1998nx, Metsaev:2000yf,
  Kallosh:1998zx}. The superstring can be defined as a non-linear
sigma model on the coset superspace
\begin{equation}
\frac{G}{H}=\frac{PSU(2,2|4)}{SO(4,1)\times SO(5)}.  \label{coset}
\end{equation}
The classical action has the Wess-Zumino-Witten form
\begin{equation}
S=-\frac{1}{2}\underset{\partial M^{3}}{\int }\mathrm{d}^{2}\sigma
\sqrt{-g} g^{\mu\nu}\left( L_{\mu}^{\hat{A}}L_{\nu}^{\hat{A}}\right)
+i\underset{M^{3}}{\int }
s^{IJ}\left( L^{\hat{A}}\wedge \overline{L}^{I}\hat{\gamma}^{\hat{A}}\wedge
L^{J}\right) ,  \label{MT}
\end{equation}
where $g^{\mu\nu}, \mu,\nu=0,1$ represents the worldsheet
metric, $\hat{A
}=(A,A^{\prime })$ with $A=\left( 0,...,4\right) $ and $A^{\prime }=\left(
5,...,9\right) $ denotes the tangent space indices for $AdS_{5}$ and $
S^{5}$ respectively,
$s^{IJ}=\mathrm{diag}(1,-1),I,J=\left(1,2\right)$;
$\hat{\gamma}^{A}\equiv \gamma ^{A}$, $\hat{\gamma}^{A^{\prime
}}\equiv i\gamma^{A^{^{\prime }}}$ ($\gamma^{(A,A')}$ represent ten
dimensional Dirac matrices),
$s^{IJ}=\mathrm{diag}(1,-1),I,J=\left( 1,2\right)$.  We use the
convention  that repeated indices are summed.
The supervielbeins $L^{\hat{A}}$ and $L^{I}$ are defined by the
left-invariant Cartan 1-forms in the $so(4,1)\oplus so(5)$ basis of
$psu(2,2|4)$ (lower case letters denote Lie algebras) as follows
\begin{eqnarray}
G^{-1}\mathrm{d}G &=& \hat{L}^{A} \hat{P}_{A}+L^{A^{\prime
}}P_{A^{\prime }}+\frac{1}{2}
 \hat{L}^{AB} \hat{J}_{AB}+\frac{1}{2}L^{A^{\prime }B^{\prime }}J_{A^{\prime
}B^{\prime }}+L^{\alpha \alpha ^{\prime }I}Q_{\alpha \alpha ^{\prime
   }I}, \label{cartanform1}
\end{eqnarray}
where $\alpha,\alpha' = 1,2,\cdots ,32$  $L^{\hat{A}} =
\left(\hat{L}^{A}, L^{A'}\right)$ and with
\begin{equation}
L^{\hat{A}} = \mathrm{d}X^{M}L_{M}^{\hat{A}},  \label{cartan1-forms}
\end{equation}
and $X^{M} =\left( x,\theta ^{I}\right)$ represent the
bosonic and fermionic string coordinates in the target space.

By solving the Cartan-Maurer equation (zero curvature condition) one
can explicitly determine
\begin{align}
L^{I}& =\left( \left( \frac{\sinh \mathcal{M}}{\mathcal{M}}\right) D\theta
\right) ^{I},  \notag \\
L^{\hat{A}}& =e^{\hat{A}}(x)\mathrm{d}x -i\overline{
\theta }\gamma ^{\hat{A}}\left( \left( \frac{\sinh \mathcal{M}/2}{\mathcal{M}
/2}\right) ^{2}D\theta \right) ,  \label{supervielbeins}
\end{align}
where
\begin{align}
\left( \mathcal{M}^{2}\right) ^{IJ}& =\epsilon ^{IK}\left( -\gamma
^{A}\theta ^{K}\bar{\theta}^{J}\gamma ^{A}+\gamma ^{A^{\prime }}\theta ^{K}
\bar{\theta}^{J}\gamma ^{A^{\prime }}\right)  \notag \\
& \qquad +\frac{1}{2}\epsilon ^{KJ}\left( \gamma ^{AB}\theta ^{I}\bar{\theta}
^{K}\gamma ^{AB}-\gamma ^{A^{\prime }B^{\prime }}\theta ^{I}\bar{\theta}
^{K}\gamma ^{A^{\prime }B^{\prime }}\right) .  \label{M}
\end{align}
Denoting by $\left( e^{\hat{A}},\omega
^{\hat{A}\hat{B}}\right) $  the bosonic vielbein and the spin connection
respectively, the covariant differential of the fermions is given by
\begin{equation}
\left( D\theta \right) ^{I}=\left[ \delta ^{IJ}\left( \mathrm{d}+\frac{1}{4}
\ \omega ^{\hat{A}\hat{B}}\gamma _{\hat{A}\hat{B}}\right) -\frac{i}{2}
\epsilon ^{IJ}\ e^{\hat{A}}\gamma _{\hat{A}}\right] \theta ^{J}.
\label{covdif}
\end{equation}

The equations of motion following from action (\ref{MT}) take the forms
\begin{align}
\sqrt{-g}g^{\mu \nu}\left( \nabla
_{\mu}\hat{L}_{\nu}^{A}+\hat{L}_{\mu}^{AB}\hat{L}_{\nu}^{B}\right)
+i\epsilon ^{\mu \nu}s^{IJ}\bar{L}_{\mu}^{I}\gamma ^{A}L_{\nu}^{J}& =0,
\label{eqofm1} \\
&  \notag \\
\sqrt{-g}g^{\mu \nu}\left( \nabla _{\mu}L_{\nu}^{A^{\prime
}}+L_{\mu}^{A^{\prime }B^{\prime }}L_{\nu}^{B^{\prime }}\right)
-\epsilon ^{\mu \nu}s^{IJ}\bar{L} _{\mu}^{I}\gamma ^{A^{\prime
}}L_{\nu}^{J}& =0,  \label{eqofm2} \\
&  \notag \\
\left( \gamma ^{A}L_{\mu}^{A}+i\gamma ^{A^{\prime }}L_{\mu}^{A^{\prime
}}\right) \left( \sqrt{-g}g^{\mu \nu}\delta ^{IJ}-\epsilon ^{\mu
  \nu}s^{IJ}\right) L_{\nu}^{J}& =0,
\label{eqofm3}
\end{align}
with $\nabla _{\mu}$ representing the covariant derivative on the worldsheet.

To consider the integrability properties of the sigma model we will need
some of the properties of the superalgebra $psu(2,2|4)$ which we
briefly review  in  Appendix {\bf A}.
Let us consider the map $G$ from the string worldsheet into the graded group
$PSU(2,2|4)$. In this case, the current $1$-form $J=-G^{-1}\mathrm{d}G$
belongs to the superalgebra and, therefore, can be decomposed as
\begin{equation}
J=-G^{-1}\mathrm{d}G = H+P+Q^{1}+Q^{2},  \label{curdecomp1}
\end{equation}
where, using the notations in Appendix {\bf A}, we can identify
\begin{equation}
H=H_{0},\quad Q^{1}=H_{1},\quad P=H_{2},\quad Q^{2}=H_{3}.  \label{ident}
\end{equation}
In terms of $L^{\hat{A}}$, $L^{\hat{A}\hat{B}}$ and $L^{I}$ one can also read
off the following expressions using (\ref{cartanform1}):
\begin{eqnarray}
H &=&\frac{1}{2} \hat{L}^{AB} \hat{J}_{AB}+\frac{1}{2}L^{A' B'}J_{A' B'},  \notag \\
&&  \notag \\
P &=&\hat{L}^{A}\hat{P}_{A}+L^{A^{\prime }}P_{A^{\prime }},  \label{identify2}
\\
&&  \notag \\
Q^{I} &=&L^{\alpha \alpha ^{\prime }I}Q_{\alpha \alpha ^{\prime }I}.  \notag
\end{eqnarray}
From the definition of the current in (\ref{curdecomp1}), we see that it
satisfies the zero curvature condition
\begin{equation}
\mathrm{d}J-J\wedge J=0.  \label{zc}
\end{equation}
In terms of the components of the current (\ref{ident}), the equations of
motion can be written as \cite{Bena:2003wd}
\begin{align}
\mathrm{d}\ {}^{\ast }P& ={}^{\ast }P\wedge H+H\wedge {}^{\ast }P+\frac{1}{2}
\left( Q\wedge Q^{\prime }+Q^{\prime }\wedge Q\right) ,  \notag \\
&  \notag \\
0& =P\wedge ({}^{\ast }Q-Q^{\prime })+({}^{\ast }Q-Q^{\prime })\wedge P,
\label{eqom1} \\
&  \notag \\
0& =P\wedge (Q-{}^{\ast }Q^{\prime })+(Q-{}^{\ast }Q^{\prime })\wedge P,
\notag
\end{align}
where $^{\ast }$ denotes the Hodge star operation and we have defined $
Q\equiv Q^{1}+Q^{2},Q^{\prime }\equiv Q^{1}-Q^{2}$.

Classical integrability is established by writing down a one parameter
family of currents $\hat{J}(t)\equiv -\hat{G}^{-1}(t)\mathrm{d}\hat{G}(t),$
where $t$ is a constant spectral parameter, satisfying the flatness
condition~\cite{Bena:2003wd}. We will use here the convenient
form of the one parameter family of currents
presented in~\cite{Das:2004hy} (here we are suppressing the
dependence on the worldsheet coordinates of all variables)
\begin{equation}
\hat{J}(t)=H+\frac{1+t^{2}}{1-t^{2}}\ P+\frac{2t}{1-t^{2}}\ {}^{\ast }P+
\sqrt{\frac{1}{1-t^{2}}}\ Q+\sqrt{\frac{t^{2}}{1-t^{2}}}\ Q^{\prime },
\label{flatcurrent}
\end{equation}
such that $\hat{J}(t=0)=J$ of (\ref{curdecomp1}). It is easy to check that
the vanishing curvature condition for this new current
\begin{equation}
\mathrm{d}\hat{J}-\hat{J}\wedge \hat{J}=0,  \label{flatness2}
\end{equation}
leads to all the equations of motion (\ref{eqom1}) as well as the zero
curvature condition (\ref{zc}).

The sigma model action which leads to the equations of motion
(\ref{eqom1}) has the form
\begin{equation}
S=\frac{1}{2}\int \mathrm{str}\left( P\wedge {}^{\ast }P-Q^{1}\wedge {Q^{2}}
\right) ,  \label{newaction}
\end{equation}
The constraint analysis for this theory can be carried out in a
straight forward manner and the resulting Hamiltonian including the
primary constraints has the form~\cite{Das:2004hy}
\begin{equation}
\mathcal{H}_{T}=\mathrm{str}\left( \frac{1}{2}\left(
P_{0}^{2}+P_{1}^{2}\right) +\lambda _{1}\varphi _{1}+\lambda _{2}\varphi
_{2}+\lambda _{3}\varphi _{3}\right) ,  \label{hamiltonain}
\end{equation}
where $\lambda _{1},$ $\lambda _{2}$ and $\lambda _{3}$ denote the Lagrange
multipliers corresponding to the three primary constraints
\begin{eqnarray}
\varphi_{1} &
=&-\partial_{1}{\cal P}_{H}+[H_{1},{\cal P}_{H}]+[P_{1},{\cal P}]+[Q_{1}^{1}
,{\cal P}_{Q^{2}}]+[Q_{1}^{2},{\cal P}_{Q^{1}}] \approx 0,  \label{Hconstraint}
\\
\varphi_{2} &
=&-\frac{1}{2}Q_{1}^{1}-\partial_{1}{\cal P}_{Q^{1}}+[H_{1},{\cal P}_{Q^{1}
}]+[P_{1},{\cal P}_{Q^{2}}]+[Q_{1}^{1},{\cal P}_{H}]+[Q_{1}^{2},{\cal
    P}_{P}] \approx 0,\\
\varphi_{3} &
=&\frac{1}{2}Q_{1}^{2}-\partial_{1}{\cal P}_{Q^{2}}+[H_{1},{\cal P}_{Q^{2}}
]+[P_{1},{\cal P}_{Q^{1}}]+[Q_{1}^{1},{\cal P}_{P}]+[Q_{1}^{2},{\cal
    P}_{H}] \approx 0.
\end{eqnarray}
Here ${\cal P}$ denotes the canonical momenta corresponding to the
respective variables. The first constraint $\varphi _{1}$
is easily checked to be stationary under the evolution of the system
while requiring the constraints $\varphi
_{2},\varphi _{3}$ to be stationary determines two of the Lagrange
multipliers to correspond to
\begin{equation}
\lambda _{2}=-Q_{1}^{2},\quad \lambda _{3}=Q_{1}^{1}.  \label{lam23}
\end{equation}
In the conformal gauge adopted in~\cite{Das:2004hy} the three
primary constraints $\left( \varphi _{1},\varphi _{2},\varphi _{3}\right) $
should be supplemented with the standard Virasoro constraints, which
in the present notation take the forms
\begin{align}
\varphi _{4} &=\frac{1}{2}\ \mbox{str}(P_{0}^{2}+P_{1}^{2}) \approx 0,
\label{1stVirasoro} \\
\varphi _{5} &=\mbox{str}(P_{0}P_{1}) \approx 0,  \label{2ndVirasoro}
\end{align}
As explained in~\cite{Das:2004hy} it is, however, a linear
combination of these constraints, namely,

\begin{eqnarray}
\bar{\varphi}_{4} &=&\varphi _{4}+\mathrm{str}\ \left( \lambda _{2}\varphi
_{2}+\lambda _{3}\varphi _{3}\right) \approx 0,  \notag \\
&&  \label{modvirasoro} \\
\bar{\varphi}_{5} &=&\varphi _{5}-\mathrm{str}\ \left( \lambda _{2}\varphi
_{2}-\lambda _{3}\varphi _{3}\right) \approx 0,  \notag
\end{eqnarray}
which can be easily checked to correspond to first class constraints and
generate the transformations associated with the reparameterization
invariance. The two fermionic
constraints $\left( \varphi _{2},\varphi _{3}\right) $ are reducible
with half of them being first class constraints responsible for the $
\kappa $-symmetry. It is however not possible to split $\left( \varphi
_{2},\varphi _{3}\right) $ into the first and second classes in a covariant
manner. As a result, one cannot carry out the standard procedure
for determining the fundamental Dirac brackets of the
theory. Therefore, we are forced to give up covariant quantization and
consider instead a gauge fixed theory. The most
convenient and simple choice from our point of view seems to be the
light-cone gauge fixed action discussed by~\cite{Metsaev:2000yf} and
later presented in the phase-space Lagrangian formalism
in~\cite{Metsaev:2000yu}. In the next section we give only a short
description of  the key features (relevant for our work) of the
light-cone gauge fixed string on $AdS_{5}\times S^{5}$, referring the
reader to the original papers~\cite{Metsaev:2000yf}
and~\cite{Metsaev:2000yu} for details.

\section{Light-cone gauge fixed string action on $AdS_{5}\times S^{5}$}
\label{sec:lightcone}
The first step in fixing the light-cone gauge is to re-write the Cartan
1-forms, given in (\ref{cartan1-forms}) in the $so(4,1)\oplus so(5)$ basis,
in the $so(3,1)\oplus su(4)$ basis. The explicit transformation is described
in detail in~\cite{Metsaev:2000yf}. We restrict here our
discussion to the bosonic case for simplicity. In this basis, the bosonic
part of the Cartan 1-form takes the form
\begin{equation}
\left( G^{-1}\mathrm{d}G\right) _{\rm
  bosonic}=L_{P}^{a}P_{a}+L_{K}^{a}K_{a}+L_{D}D+ \frac{
1}{2}L^{ab}J_{ab}+L_{j}^{i}J_{i}^{j},  \label{cartanlightcone}
\end{equation}
where the index $A$ splits into $(a,4)$ and the new generators
correspond to those for translations, conformal boosts, dilatation
\begin{equation}
P^{a}=\hat{P}^{a}+\hat{J}^{4a},\quad K^{a}=\frac{1}{2}\left( -\hat{P}^{a}+
\hat{J}^{4a}\right),\quad D=-\hat{P}^{4},\quad J^{ab}=\hat{J}^{ab},
\label{conformalalge}
\end{equation}
and the $su(4)$ generators
\begin{equation}
J_{j}^{i}=-\frac{i}{2}\left( \gamma ^{A^{\prime }}\right)
_{j}^{i}P^{A^{\prime }}+\frac{1}{4}\left( \gamma ^{A^{\prime }B^{\prime
}}\right) _{j}^{i}J^{A^{\prime }B^{\prime }}.  \label{su4gen}
\end{equation}
In Appendix {\bf B} we list some relations between the old and the new
basis.
These bosonic generators form the $so(4,2)\oplus so(6)$ subalgebra of $
psu(2,2|4)$. In the light-cone coordinates
\begin{eqnarray}
x^{a} &=&\left( x^{+},x^{-},x,\bar{x}\right),  \notag \\
&&  \notag \\
x^{\pm } &\equiv &\frac{1}{\sqrt{2}}\left( x^{3}\pm x^{0}\right),
\label{lc_coordinates} \\
&&  \notag \\
(x,\bar{x}) &\equiv &\frac{1}{\sqrt{2}}\left( x^{1}\pm ix^{2}\right),
\notag
\end{eqnarray}
the above generators $\left( P^{a},J^{ab},K^{a}\right) $ split as
\begin{equation}
\left( P^{\pm },P\equiv P^{x},\bar{P}\equiv P^{\bar{x}},J^{\pm x},J^{\pm 
\bar{x}},J^{+-},J^{x\bar{x}},K\equiv K^{x},K^{\pm },\bar{K}\equiv K^{\bar{x}
}\right).  \label{split}
\end{equation}
Since the metric in the light-cone coordinates is not diagonal, the
scalar product takes the form
\begin{equation}
A^{a}B_{a}=A^{+}B^{-}+A^{-}B^{+}+A^{x}B^{\bar{x}}+A^{\bar{x}}B^{x}.
\label{sumrule}
\end{equation}
The fermionic generators in these coordinates split into two sets of
supercharges  - regular and conformal
\begin{equation}
Q^{\pm i},Q_{i}^{\pm },S^{\pm i},S_{i}^{\pm }.  \label{supercharges2}
\end{equation}

Finally, to fix the $\kappa -$symmetry one chooses a specific representative
of the supercoset $\frac{PSU(2,2|4)}{SO(4,1)\times SO(5)}$ parametrized by
the coordinates $\left( x^{a},y_{j}^{i},\phi ,\theta ^{\pm
  i},\theta^{\pm}_{i}, \eta ^{\pm i},\eta^{\pm}_{i}\right) $ in the
following form:
\begin{equation}
G=G_{x,\theta }G_{\eta }G_{y}G_{\phi},  \label{repres1}
\end{equation}
where
\begin{eqnarray}
G_{x,\theta } &=&\exp \left( x^{a}P_{a}+\theta ^{-i}Q_{i}^{+}+\theta
^{+i}Q_{i}^{-}+\theta _{i}^{-}Q^{+i}+\theta _{i}^{+}Q^{-i}\right),
\notag \\
&&  \notag \\
G_{\eta } &=&\exp \left( \eta ^{-i}S_{i}^{+}+\eta ^{+i}S_{i}^{-}+\eta
_{i}^{-}S^{+i}+\eta _{i}^{+}S^{-i}\right),  \notag \\
&&  \label{repres2} \\
G_{y} &=&\exp \left( y_{j}^{i}J_{i}^{j}\right),  \notag \\
&&  \notag \\
G_{\phi } &=&\exp \left( \phi D\right),  \notag
\end{eqnarray}
where the $S^{5}$ coordinates $y^{A^{\prime }}$ appear through
$y_{j}^{i}\equiv \frac{i}{2}\left( \gamma ^{A^{\prime }}\right)
_{j}^{i}$ $y^{A^{\prime }}$.
Analogous to the flat space case, the $\kappa$ symmetry is then
fixed by setting the positive components of the spinors to zero
\begin{equation}
\theta ^{+i}=\theta _{i}^{+}=\eta ^{+i}=\eta _{i}^{+}=0.
\label{kappafixed}
\end{equation}
The coset space representative (\ref{repres2}) will, as a result, take a
simpler form, and the corresponding Cartan 1-forms can be written down in
the explicit form (see also the next section). The Cartan 1-forms associated
with the bosonic sector are given by~\cite{Metsaev:2000yf}:
\begin{eqnarray}
L_{P}^{+} &=&e^{\phi }\mathrm{d}x^{+},\hspace{0.23in}L_{P}^{-}=e^{\phi
}\left(\mathrm{d} x^{-}-
\frac{i}{2}\widetilde{\theta }^{i}\mathrm{d}\widetilde{\theta
}_{i}-\frac{i}{2}
\widetilde{\theta }_{i}\mathrm{d}\widetilde{\theta }^{i}\right),  \notag \\
&&  \notag \\
L_{P}^{x} &=&e^{\phi
}\mathrm{d}x,\hspace{0.23in}L_{P}^{\bar{x}}=e^{\phi }\mathrm{d}\bar{x},
\hspace{0.23in}L_{D}=\mathrm{d}\phi,  \label{bosonic_cartan} \\
&&  \notag \\
L_{j}^{i} &=&\left(\mathrm{d}UU^{-1}\right) _{j}^{i}+i\left(
\widetilde{\eta }^{i}
\widetilde{\eta }_{j}-\frac{1}{4}\widetilde{\eta }^{2}\delta _{j}^{i}\right)
\mathrm{d}x^{+},  \notag
\end{eqnarray}
while the fermionic ones have the following form:
\begin{eqnarray}
L_{K}^{-} &=&e^{-\phi }\left( \frac{1}{4}\left( \widetilde{\eta }^{2}\right)
^{2}\mathrm{d}x^{+}+\frac{i}{2}\widetilde{\eta
}^{i}\mathrm{d}\widetilde{\eta }_{i}+
\frac{i}{2}\widetilde{\eta }_{i}\mathrm{d}\widetilde{\eta
}^{i}\right),   \notag \\
&&  \notag \\
L_{Q}^{-i} &=&e^{\frac{\phi }{2}}\left(\mathrm{d}\widetilde{\theta
}^{i}+i\widetilde{\eta
}^{i}\mathrm{d}x\right),\hspace{0.23in}L_{Q_i}^{-}=e^{\frac{\phi
  }{2}}\left(\mathrm{d}\widetilde{\theta}_{i}-i\widetilde{\eta
}_{i}\mathrm{d}\bar{x}\right),  \notag \\
&&  \label{fermionic_cartan} \\
L_{Q}^{+i} &=&-ie^{\frac{\phi }{2}}\widetilde{\eta
}^{i}\mathrm{d}x^{+}, \hspace{0.23in}L_{Q_i}^{+}=ie^{\frac{\phi
  }{2}}\widetilde{\eta }_{i}\mathrm{d}x^{+},  \notag \\
&&  \notag \\
L_{S}^{-i} &=&e^{-\frac{\phi }{2}}\left(\mathrm{d}\widetilde{\eta
}^{i}+\frac{i}{2}
\widetilde{\eta }^{2}\widetilde{\eta
}^{i}\mathrm{d}x^{+}\right),\hspace{0.23in}
L_{S_i}^{-}=e^{-\frac{\phi }{2}}\left(\mathrm{d}\widetilde{\eta
}_{i}-\frac{i}{2}
\widetilde{\eta }^{2}\widetilde{\eta }_{i}\mathrm{d}x^{+}\right),  \notag
\end{eqnarray}
where for simplicity we have defined
\begin{eqnarray}
\theta ^{i} &\equiv &\theta ^{-i},\hspace{0.13in}\eta ^{i}\equiv \eta
^{-i},
\hspace{0.13in}\theta _{i}\equiv \theta _{i}^{-},\hspace{0.13in}\eta
_{i}\equiv \eta _{i}^{-},  \notag \\
&&  \label{thetadual} \\
\widetilde{\theta }^{i} &\equiv &U_{j}^{i}\theta ^{j},\hspace{0.13in}
\widetilde{\theta }_{i}\equiv \theta _{j}(U^{-1})_{i}^{j}.  \notag
\end{eqnarray}
The matrix $U_{j}^{i}$ has the following explicit dependence on coordinates $
u^{M}$ $(M=1...6)$ parameterizing \footnote{
The metric on $AdS_{5}\times S^{5}$ is chosen to be of the form
$\mathrm{d}s^{2}=e^{2\phi
}\mathrm{d}x^{A}\mathrm{d}x^{A}+\left(\mathrm{d}\phi\right)
^{2}+\mathrm{d}u^{M} \mathrm{d}u^{M},
\hspace{0.13in}u^{M}u^{M}=1$} the sphere $S^{5}$
\begin{eqnarray}
U_{j}^{i} &=&\frac{\rho ^{6ik}\left( \rho _{kj}^{6}+\rho
_{kj}^{M}u^{M}\right) }{\sqrt{2}\sqrt{1+u^{6}}},  \notag \\
&&  \label{Umatrix} \\
\left( U^{-1}\right) _{j}^{i} &=&\frac{\left( \rho ^{6ik}+\rho
^{Mik}u^{M}\right) \rho _{kj}^{6}}{\sqrt{2}\sqrt{1+u^{6}}}.  \notag
\end{eqnarray}
Here $\rho _{kj}^{M}$ are matrices satisfying
\begin{equation}
\rho ^{M}\overline{\rho }^{N}+\rho ^{N}\overline{\rho }^{M}=2\delta
^{MN}, \label{roalgebra}
\end{equation}
\begin{eqnarray}
\rho ^{M} &=&\left( \rho ^{M}\right) ^{ij},\quad \overline{\rho
}^{M}=\left( \rho^{M}\right) _{ij},  \notag \\
&&  \notag \\
\hat{\rho}^{MN} &=&\frac{1}{2}\left( \overline{\rho }^{M}\rho ^{N}-\overline{
\rho }^{N}\rho ^{M}\right),  \label{ronotat} \\
&&  \notag \\
\rho ^{MN} &=&\frac{1}{2}\left( \rho ^{M}\overline{\rho }^{N}-\rho ^{N}
\overline{\rho }^{M}\right).  \notag
\end{eqnarray}
The gauge-fixed Lagrangian density, in these coordinates and
notations, takes the form
\begin{eqnarray}
& &\mathcal{L}=-\sqrt{-g}g^{\mu \nu }\left( e^{2\phi }\left( \partial _{\mu
}x^{+}\partial _{\nu }x^{-}+\partial _{\mu }x\partial _{\nu }\bar{x}\right) +
\frac{1}{2}\partial _{\mu }\phi \partial _{\nu }\phi +\frac{1}{2}D_{\mu
}u^{M}D_{\nu }u^{M}\right)  \notag \\
&&  \notag \\
&&\; -\frac{i}{2}\sqrt{-g}g^{\mu \nu }e^{2\phi } \partial _{\mu
}x^{+}\left(\theta ^{i}\partial _{\nu }\theta _{i}+\eta ^{i}\partial
_{\nu }\eta_{i}+\theta _{i}\partial _{\nu }\theta ^{i}+\eta
_{i}\partial _{\nu }\eta^{i}+ie^{2\phi }\partial _{\nu }x^{+}\left(
\eta ^{2}\right) ^{2}\right) \notag \\
&&  \notag \\
&&\; +\epsilon ^{\mu \nu }e^{2\phi }\partial _{\mu }x^{+}\eta ^{i}\left( \rho
^{M}\right) _{ij}u^{M}\left( \partial _{\nu }\theta ^{j}-i\sqrt{2}e^{\phi
}\eta ^{j}\partial _{\nu }x\right) + {\rm h.c.},  \label{gaugedfixedL}
\end{eqnarray}
where
\begin{equation}
D_{\mu }u^{M}\equiv \partial _{\mu }u^{M}+ie^{2\phi }\eta _{i}\left( \rho
^{MN}\right) _{j}^{i}\eta ^{j}u^{N}\partial _{\mu }x^{+}.  \label{covderiv}
\end{equation}

The next step is to derive the set of constraints as well as the
Hamiltonian.
This analysis is best carried out in the (mixed) phase-space
Lagrangian formalism~\cite{Metsaev:2000yu}. The advantage of this
method is that it allows us to choose the conventional bosonic gauge
\begin{equation}
x^{+}=\tau ,\quad \mathcal{P}^{+}= {\rm const},  \label{bosonicgauge}
\end{equation}
where $\mathcal{P}^{+}$ is the conjugate momentum for $x^{-}$. In the flat
space this translates into the usual light-cone gauge fixing in the
coordinate space which is not generally consistent in a curved
background~\cite{Horowitz:1990sr, Rudd:1994ss}. For the present case of the $
AdS_{5}\times S^{5}$ background, imposing the condition
(\ref{bosonicgauge}) in the phase space is equivalent in the
coordinate space to identifying
\begin{equation}
h^{00}=-p^{+},\quad h^{11}=\left( p^{+}\right) ^{-1}e^{4\phi },\quad 
h^{01}=h^{10}=0,  \label{Virasorofixed}
\end{equation}
where
\begin{equation}
h^{\mu \nu }\equiv \sqrt{-g}g^{\mu \nu }e^{2\phi}.  \label{newmetric}
\end{equation}
Another advantage of the phase-space Lagrangian is that one can easily
perform explicit integration over some of the degrees of freedom, avoiding
subtleties~\cite{Rudd:1994ss} associated with the ghost fields.

Even after the reparametrization and $\kappa$-symmetries are fixed, there
remain some residual fermionic and bosonic constraints (e.g. $u^{M}u^{M}=1)$
and it is necessary to evaluate the Dirac brackets taking into account
these and the corresponding brackets have already been obtained
in~\cite{Metsaev:2000yu}. The Hamiltonian density has the following form:
\begin{eqnarray}
\mathcal{H} &=&\frac{1}{2p^{+}}\left[ 2\mathcal{P}\mathcal{\bar{P}}+2e^{4\phi }
x^{'}\bar{x}^{'} \right.  \notag \\
&&\left. +e^{2\phi }\left( \mathcal{P}_{\phi }^{2}+ (\phi')^{2}+\mathcal{
P}^{M}\mathcal{P}^{M}+ u^{'\ M} u^{'\ M}+p^{+2}\left( \eta
^{2}\right) ^{2}+4p^{+}\eta _{i}l_{j}^{i}\eta ^{j}\right) \right]  \notag \\
&&  \label{Hgaugedfixed} \\
&&-e^{2\phi }\eta ^{i}y_{ij}\left(\theta^{'\ j}-i\sqrt{2}e^{\phi }\eta
^{j} x'\right) -e^{2\phi }\eta _{i}y^{ij}\left(\theta'_{j}+i
\sqrt{2}e^{\phi }\eta _{j} x'\right),  \notag
\end{eqnarray}
where a ``prime'' denotes a derivative with respect to $\sigma$ and
\begin{eqnarray}
y_{ij} &\equiv &\left( \rho ^{M}\right) _{ij}u^{M},  \notag \\
&&  \label{lij} \\
l_{j}^{i} &\equiv &\frac{i}{2}\left( \rho ^{MN}\right) _{j}^{i}u^{M}\mathcal{
P}^{N}.  \notag
\end{eqnarray}
The fundamental Dirac brackets can be determined to be
\begin{eqnarray}
\{ \mathcal{P}(\sigma ),\bar{x}(\sigma ^{\prime })\}_{D} &=&\delta
(\sigma -\sigma ^{\prime }) = \{\mathcal{\bar{P}}(\sigma
),x(\sigma ^{\prime })\}_{D} = \{ \mathcal{P}_{\phi }(\sigma
  ),\phi (\sigma ^{\prime })\}_{D},\notag \\
&&  \notag \\
\{ \mathcal{P}^{M}(\sigma
),u^{N}(\sigma ^{\prime })\}_{D} &= &\left( \delta
  ^{MN}-u^{M}u^{N}\right) \delta
(\sigma -\sigma ^{\prime }),  \notag \\
&&  \notag \\
\{ \mathcal{P}^{M}(\sigma ),\mathcal{P}^{N}(\sigma ^{\prime })\}_{D}
&=&\left( u^{M}\mathcal{P}^{N}-u^{N}\mathcal{P}^{M}\right) \delta (\sigma
-\sigma ^{\prime }),  \label{Dirac_central} \\
&&  \notag \\
\{ \theta _{i}(\sigma ),\theta ^{j}(\sigma ^{\prime })\}_{D} &=&\frac{i}{
p^{+}}\delta _{i}^{j}\delta (\sigma -\sigma ^{\prime }), \notag \\
&&  \notag \\
\{ x_{0}^{-},\theta _{i}(\sigma)\}_{D} &=&\frac{1}{
2p^{+}}\theta _{i}(\sigma),\quad \{
  x_{0}^{-},\theta^{i}(\sigma)\}_{D}=\frac{1}{2p^{+}}\theta ^{i}(\sigma).   \notag
\end{eqnarray}
Here $x_{0}^{-}$ is the zero mode of the $x^{-}$coordinate. The
other brackets
can be obtained from (\ref{Dirac_central}) by replacing $\left( \theta
_{i}\rightarrow \eta _{i};\theta ^{i}\rightarrow \eta ^{i}\right)$. We note
here that the constraint $u^{M}u^{M}=1$ leads to the secondary constraint $
\mathcal{P}^{M}u^{M}=0$ (both are second class constraints) which
play a crucial role in studying the integrability properties of the
model. Some related interesting and explicit calculations have
recently been presented in~\cite{Alday:2005gi}.

\section{Algebra of currents without the spectral parameter}
\label{sec:algebraofcurrents}

In this section we would like to investigate the classical algebra of
the currents with the Virasoro constraints (reflecting
reparameterization invariance) and the $\kappa$-symmetry fixed. It is
not {\em a priori} obvious what would result in the light-cone gauge
and  such an algebra is desirable if one is to obtain
the classical $r$-matrix. In~\cite{Das:2004hy} it was
shown  that (at least in the bosonic sector, when one
ignores the difficulties associated with the $\kappa$-symmetry) the algebra
is closed. However, the Virasoro constraints were not
completely fixed there and the algebra, although closed, contained
undesirable Schwinger terms. As pointed out in the introduction, the
presence of such terms in the algebra of the
flat currents leads to ambiguities in the computation of brackets between
the transition matrices. This, in turn, prevents one from determining
the classical $r$-matrix of the theory. Therefore, one needs to regularize such
terms. There already exist several regularization
schemes~\cite{Maillet:1985ek, Duncan:1989vg, Faddeev:1985qu}. The most
intriguing (and interesting) possibility has been
outlined in the original work by Faddeev and
Reshetikhin~\cite{Faddeev:1985qu} where one
introduces a reduced model ``by hand", equivalent to that of the two
interacting spins, and only in the end a relation between the quantum
version of the reduced model and the original classical system (with the use
of the Schwinger theorem) is argued in a limiting manner. This
reduced system is still an interesting model to study on its own, and may
turn out to be quite important after one fully understands the
limiting procedure. It
is quite remarkable that the $FR$ model has resurfaced in the context of
the $AdS$ string~\cite{Kazakov:2004qf}, namely, the string in
$AdS_{1}\times S^{3}$ background
can be written exactly as the $FR$ model of the two
interacting spins. The
key point of this construction is to recognize that the Virasoro constraints
in this model coincide exactly with the ones introduced by Faddeev and
Reshetikhin to reduce the original system (of the principal $SU(2)$
chiral model). In other words, there is no need
to impose any condition ``by hand" since the Virasoro conditions do
exactly that. Motivated by these observations, it is
important to analyze the effects of the Virasoro constraints when the
string moves on the $
AdS_{5}\times S^{5}$ background. The ultimate goal is to elucidate whether
fixing completely the reparametrization invariance will lead to some sort of
well-defined spin system as it happens in the $AdS_{1}\times S^{3}$
background.

Before we proceed, let us note that for simplicity we will employ an
index free tensor notation following \cite{Faddeev:1987ph} defined as
\begin{equation}
\overset{1}{A}=A\otimes I,\text{ \ }\overset{2}{B}=I\otimes B,  \label{a1b2}
\end{equation}
where
\begin{equation}
(A\otimes B)_{ij,km}=A_{ik}B_{jm}.  \label{abtensor}
\end{equation}
In these notations, our goal is to calculate $\{\overset{\left( 1\right) }{J}
_{1}(t_{1,}\sigma ),\overset{\left( 2\right) }{J}_{1}(t_{2,}\sigma ^{\prime
})\}_{D}$, where $t_{i}$ is the spectral parameter, after the
reparametrization and the $
\kappa$-symmetry have been fixed as described in the previous
section. To avoid
technical complications and to illustrate the essential points, we restrict
ourselves in this paper to the bosonic case and will present the full
supersymmetric case in a separate publication. Let us collect the set of the
light-cone generators as follows:
\begin{equation}
T_{a}=-\left({P}^{-},{P}^{+},{P}^{\overline{x}
},{P}^{x},{D},{J}_{j}^{i}\right),\quad a=1,2,\cdots ,6.  \label{genset}
\end{equation}
Then, one can write the components of the current in the following form:
\begin{eqnarray}
J_{0} &\equiv &A^{a}{T}_{a},  \label{J0-lc} \\
&&  \notag \\
J_{1} &\equiv &B^{a}{T}_{a}.  \label{J1-lc}
\end{eqnarray}
The explicit form of the coefficient functions $A^{a}$ and $B^{a}$
can be determined  using
the Cartan 1-forms in the light-cone fixed gauge (see previous section). We
collect below the forms of $A^{a}$ and $B^{a}$  which can
be written as functions of the coordinates and momenta only:

\begin{eqnarray}
A^{1} &=&e^{\phi},  \notag \\
&&  \notag \\
A^{2} &=&-\frac{e^{\phi }}{2\left( p^{+}\right) ^{2}}\left[ 2{\cal
    P}\overline{\cal P} +2e^{4\phi }x'\overline{x}'+e^{2\phi }\left(
  {\cal P}_{\phi }^{2}+ (\phi')^{2}+{\cal P}_{M}^{2}+u^{'\ M} u^{'\
    M}\right) \right],  \notag\\
&&  \label{Ai} \\
A^{3} &=&\frac{e^{\phi }}{p^{+}}{\cal
  P},\hspace{0.23in}A^{4}=\frac{e^{\phi }}{p^{+}}\overline{\cal P},
\hspace{0.23in}A^{5}=\frac{e^{2\phi }}{p^{+}}{\cal P}_{\phi },  \notag
\\
&&  \notag \\
A^{6} &=&\frac{e^{2\phi }}{2\left( 1+u^{6}\right) p^{+}}\left[ \rho ^{6}
\widehat{\rho }^{MN}\overline{\rho }^{6}{\cal P}^{M}u^{N}-\rho
^{A}\overline{\rho}^{6}{\cal P}^{A}\right],  \notag
\end{eqnarray}
and
\begin{eqnarray}
B^{1} &=&0,  \notag \\
&&  \notag \\
B^{2} &=&-\frac{e^{\phi }}{p^{+}}\left[ {\cal
    P}\overline{x}'+\overline{\cal P} x'+{\cal P}_{\phi }\phi'+{\cal
    P}_{M} u^{'\ M}\right],  \notag \\
&&  \label{Bj} \\
B^{3} &=&e^{\phi }x',\hspace{0.23in}B^{4}=e^{\phi
}\overline{x}',\hspace{0.23in} B^{5}=\phi',  \notag \\
&&  \notag \\
B^{6} &=&\frac{1}{2\left( 1+u^{6}\right) }\left[ \rho ^{6}\widehat{\rho }
^{MN}\overline{\rho }^{6} u^{'\ M}u^{N}-\rho ^{A}\overline{\rho }^{6}
u^{'\ A}\right],  \notag
\end{eqnarray}
where a ``prime'' denotes differentiation with respect to $\sigma$.

Using various relations listed in Appendix {\bf B} one finds that the
flat  current (\ref{flatcurrent}) $\hat{J}_{1}(\sigma ,t)$ can be
written as

\begin{equation}
\hat{J}_{1}(t)=H_{1}+\frac{1+t^{2}}{1-t^{2}}P_{1}+\frac{2t}{1-t^{2}}
    {}^{\ast} P_{1},  \label{flatcurrent-lc}
\end{equation}
where\footnote{Note that the Hodge dual is defined as ${}^{\ast}
P_{1}=-p^{+}e^{-2\phi }P_{0}$}
\begin{eqnarray}
H_{1} &=&B^{2}(\frac{1}{2}P^{+}+K^{+})+B^{3}(\frac{1}{2}\bar{P}+\bar{K}
)+B^{4}(\frac{1}{2}P+K)  \notag \\
&&+B_{\;\;\;j}^{6i}\frac{1}{2}(J_{\;\;i}^{j}+\rho ^{6jk}\rho
_{il}^{6}J_{\;\;k}^{l}),  \notag \\
&&  \notag \\
P_{1} &=&B^{2}(\frac{1}{2}P^{+}-K^{+})+B^{3}(\frac{1}{2}\bar{P}-\bar{K}
)+B^{4}(\frac{1}{2}P-K)+B^{5}D  \notag \\
&&+B_{\;\;\;j}^{6i}\frac{1}{2}(J_{\;\;i}^{j}-\rho ^{6jk}\rho
_{il}^{6}J_{\;\;k}^{l}),  \notag \\
&&  \label{h1p1p0} \\
{}^{\ast} P_{1} &=&-p^{+}e^{-2\phi }\left( A^{1}(\frac{1}{2}P^{-}-K^{-})+A^{2}(
\frac{1}{2}P^{+}-K^{+})+A^{3}(\frac{1}{2}\bar{P}-\bar{K})\right.  \notag \\
&&\left. +A^{4}(\frac{1}{2}P-K)+A^{5}D+A_{\;\;\;j}^{6i}\frac{1}{2}
(J_{\;\;i}^{j}-\rho ^{6jk}\rho _{il}^{6}J_{\;\;k}^{l})\right).  \notag
\end{eqnarray}

We can already see that the calculation of the algebra of currents
(with the spectral parameter) is
rather complicated and one requires the knowledge of the brackets
(without the spectral parameter)
\begin{eqnarray}
\{\overset{\left( 1\right) }{J}_{0}(\sigma ),\overset{\left( 2\right) }{J}
_{0}(\sigma ^{\prime })\}_{D} &=&\{A^{a},A^{b}\}_{D} T_{a}\otimes
T_{b}, \notag \\
&&  \notag \\
\{\overset{\left( 1\right) }{J}_{0}(\sigma ),\overset{\left( 2\right) }{J}
_{1}(\sigma ^{\prime })\}_{D} &=&\{A^{a},B^{b}\}_{D} T_{a}\otimes T_{b},
\label{tensor} \\
&&  \notag \\
\{\overset{\left( 1\right) }{J}_{1}(\sigma ),\overset{\left( 2\right) }{J}
_{1}(\sigma ^{\prime })\}_{D} &=&\{B^{a},B^{b}\}_{D} T_{a}\otimes
T_{b}. \notag
\end{eqnarray}
The calculation of $\{A^{a},A^{b}\}_{D}$, $\{B^{a},B^{b}\}_{D}$ and $
\{A^{a},B^{b}\}_{D}$ is in itself quite tedious. But it is even more
challenging  to
express the brackets in terms of $A^{a}$ and $B^{a}$ and only then
 the algebra of the flat currents may have a closed
form. This turns out to be a non-trivial problem itself. For instance, as it
is seen from the above table, the $S^{5}$ coordinates do not enter
covariantly in the expressions for $A^{6}$ and $B^{6}$, namely $u^{6}$
appears separately in the denominator. Nevertheless, we have
calculated and found that all the brackets, $
\{B^{a},B^{j}\}_{D}$, $\{A^{a},B^{b}\}_{D}$ and $\{A^{a},A^{b}\}_{D}$,
 can be
written as functions $F(A^{a},B^{a})$ of $A^{a}$ and $B^{a}.$The
simplest case is $\{B^{a},B^{b}\}_{D}$ where the nontrivial brackets
are given by\footnote{
To avoid cluttered notations, almost everywhere in brackets
$\{F,G\}_{D}$ we
omit manifest dependence on the word-sheet coordinate, and assume that
the
function on the left $F=F(\sigma ),$ and the function on the right $
G=G(\sigma ^{\prime }),$ i.e. $\{F,G\}_{D}\equiv \{F(\sigma
),G(\sigma^{\prime})\}_{D}$}

\begin{eqnarray}
\{B^{2},B^{3}\} &=&\frac{e^{\phi (\sigma )}}{p^{+}}B^{3}(\sigma )\partial
_{\sigma }\delta (\sigma -\sigma ^{\prime }),  \notag \\
&&  \notag \\
\{B^{2},B^{4}\} &=&\frac{e^{\phi (\sigma )}}{p^{+}}B^{4}(\sigma )\partial
_{\sigma }\delta (\sigma -\sigma ^{\prime }),  \notag \\
&&  \notag \\
\{B^{2},B^{5}\} &=&\frac{e^{\phi (\sigma )}}{p^{+}}B^{5}(\sigma )\partial
_{\sigma }\delta (\sigma -\sigma ^{\prime }),  \label{bibj} \\
&&  \notag \\
\{B^{2},B^{6}\} &=&\frac{e^{\phi (\sigma )}}{p^{+}}B^{6}(\sigma )\partial
_{\sigma }\delta (\sigma -\sigma ^{\prime }),  \notag \\
&&  \notag \\
\{B^{2},B^{2}\} &=&\frac{e^{\phi (\sigma )}}{p^{+}}B^{2}(\sigma )\partial
_{\sigma }\delta (\sigma -\sigma ^{\prime })-\frac{e^{\phi (\sigma ^{\prime
})}}{p^{+}}B^{2}(\sigma ^{\prime })\partial _{\sigma ^{\prime }}\delta
(\sigma -\sigma ^{\prime }).  \notag
\end{eqnarray}

Using this result, one can show that $\{\overset{1}{J}_{1}(\sigma ),
\overset{2}{J}_{1}(\sigma ^{\prime })\}_{D}$ can be elegantly written in the
following form:
\begin{equation}
\{\overset{1}{J}_{1}(\sigma ),\overset{2}{J}_{1}(\sigma ^{\prime
})\}_{D} =\left(
\frac{e^{\phi (\sigma )}}{p^{+}}{P}^{+}\otimes J_{1}(\sigma )+\frac{
e^{\phi (\sigma ^{\prime })}}{p^{+}}J_{1}(\sigma ^{\prime })\otimes
{P}^{+}\right) \partial _{\sigma }\delta (\sigma -\sigma ^{\prime }).
\label{J1J1}
\end{equation}
We see that the non-ultralocal terms $\sim \partial _{\sigma }\delta
(\sigma -\sigma ^{\prime })$ appear in this algebra with a weight $e^{\phi(\sigma )}$.
As we mentioned in the introduction, this is in parallel to the case of
two-dimensional sigma models coupled to the dilaton field
$\rho=e^{\phi(\sigma)}$, that are obtained from D-dimensional gravity
via dimensional reduction. In this case, as was shown in~\cite{Korotkin:1997fi}, the
presence of the dilaton field regularizes the
ambiguities in the algebra of the transition matrices arising from the
non-ultralocal terms, provided the dilaton satisfies appropriate
boundary conditions. The origin of such a
regularization can be traced back to the non-trivial dependence of
the spectral parameter on the world-sheet coordinates and the dilaton
$\phi$.
In the present case, the spectral parameter is a constant, but in the
light-cone gauge the solution of the Virasoro constraints leads to
the field $\rho \sim e^{\phi({\sigma})}$,
where $\phi(\sigma)$ denotes the radial $AdS_5$ coordinate.
It is, therefore, plausible that the non-ultralocal terms in the
present case will not to lead to ambiguities in the algebra of the
monodromy matrices. This issue will be analyzed in detail in a
separate publication. This is in contrast with what happens if one doesn't fix
the Virasoro constraints. Let us recall that in that case the algebra
corresponding
to (\ref{J1J1}) has the following form~\cite{Das:2004hy}:
\begin{align}
& \{\overset{1}{\hat{J}_{1}}(\sigma ,t_{1}),\overset{2}{\hat{J}_{1}}(\sigma
^{\prime },t_{2})\}  \notag \\
& \;=\left( \alpha \lbrack \Omega _{P},\overset{\left( 1\right) }{\hat{J}_{1}
}(\sigma ,t_{1})]+\beta \lbrack \Omega _{P},\overset{\left( 2\right) }{\hat{J
}_{1}}(\sigma ,t_{2})]+\gamma \lbrack \Omega _{H},\overset{\left( 1\right) }{
\hat{J}_{1}}(\sigma ,t_{1})+\overset{\left( 2\right) }{\hat{J}_{1}}(\sigma
,t_{2})]\right) \delta (\sigma -\sigma ^{\prime })  \notag \\
& \qquad  \notag \\
& \qquad +\Lambda \partial _{\sigma }\delta (\sigma -\sigma ^{\prime }),
\label{central}
\end{align}
where in the second term $\Lambda =\Lambda (t_{1,}t_{2})$ is a constant.
Therefore, integrating (\ref{central}) will inevitably lead to
ambiguities~\cite{deVega:1983gy, Maillet:1985ek, Korotkin:1997fi}.

In the Appendices {\bf C} and {\bf D} we have collected the remaining
brackets  $\{A^{a},B^{b}\}_{D},$ and $\{A^{a},A^{b}\}_{D}$. Some of these
brackets are rather
complicated, however one can readily verify that all potentially dangerous
terms are multiplied by $e^{\phi (\sigma )}$ and thereby likely to
lead to unambiguous algebra for the monodromy matrices much along the
lines in~\cite{Korotkin:1997fi}.

\section{Algebra of currents with the spectral parameter}
\label{sec:algebrawithspectral}

In this section we will collect all the pieces into the calculation of
the algebra of the flat current with the spectral
parameter (\ref{flatcurrent-lc}) which leads to the monodromy
matrix. Let us redefine the basis of the generators as
\begin{eqnarray}
&&t^1_H = \frac{1}{2} P^- + K^-, \quad t^2_H = \frac{1}{2} P^+ + K^+, \quad
t^3_H = \frac{1}{2} \bar{P} + \bar{K},  \notag \\
&&t^4_H = \frac{1}{2} P + K,
\quad t^5_H =0,\quad (t^6_H)^j_i = \frac{1}{2} (J^j_i +
\rho^{jk}\rho_{il} J^l_k).
\end{eqnarray}
and
\begin{eqnarray}
&&t^1_P = \frac{1}{2} P^- - K^-, \quad t^2_P = \frac{1}{2} P^+ - K^+, \quad
t^3_P = \frac{1}{2} \bar{P} - \bar{K},   \notag \\
&&t^4_P = \frac{1}{2} P - K,
\quad t^5_P =D,\quad (t^6_P)^j_i = \frac{1}{2} (J^j_i -
\rho^{jk}\rho_{il} J^l_k).
\end{eqnarray}
The one parameter family of flat currents, in this basis, is given by
(see~\ref{flatcurrent})
\begin{equation}
\hat{J}(t)= H + a(t) P + b(t) {}^{\ast} P,
\end{equation}
where
\begin{equation}
a(t) = \frac{1+t^2}{1-t^2}, \quad b(t) = \frac{2t}{1-t^2},
\end{equation}
and the components of this can be written in terms of the generators as  
\begin{eqnarray}
&&H_0=A^at^a_H, \quad H_1 = B^a t^a_H, \quad P_0 = A^a t^a_P, \quad P_1 =B^a
t^a_P,  \notag \\
&& {}^{\ast}P_1 = -p^+e^{-2\phi} P_0 = -p^+ e^{-2\phi} A^at^a_P \equiv
\bar{A}^a t^a_P,
\end{eqnarray}
where the sum runs from a=1 to 6.

In this basis, the result of the calculation $\{\overset{1}{\hat{J}_{1}} 
(\sigma ,t_{1}),\overset{2}{\hat{J}_{1}}(\sigma ^{\prime
},t_{2})\}_{D}$ can be written as\footnote{Here we suppress the
dependence on $\sigma$ and use the notations $\phi^{\prime}\equiv
\phi(\sigma^{\prime})$, $J_{1}^{\prime}(t)\equiv J_{1}(\sigma^{\prime},t)$ }

\begin{eqnarray}
&&\{\overset{1}{\hat{J}_{1}}(\sigma ,t_{1}),\overset{2}{\hat{J}_{1}}(\sigma
^{\prime },t_{2})\}_{D}\notag \\
&=&-b(t_{1})t_{P}^{2}\otimes \lbrack \hat{J}_{0}(t_{2}),\hat{J}_{1}(t_{2})]e^{-\phi
}\delta (\sigma -\sigma ^{\prime
})+b(t_{2})[\hat{J}_{0}(t_{1}),\hat{J}_{1}(t_{1})]\otimes t_{P}^{2}e^{-\phi }\delta
(\sigma -\sigma ^{\prime }) \notag\\
&&-b(t_1) t^5_p \otimes \hat{J}_1(-t_2) \delta (\sigma -\sigma^{\prime}) +b(t_2)
\hat{J}_1(-t_1) \otimes t^5_P \delta (\sigma -\sigma^{\prime})  \notag \\
&& +t_{H}^{2}\otimes \hat{J}_{1}(t_{2})\frac{e^{\phi }}{p^{+}}\partial \delta
(\sigma -\sigma ^{\prime })+\hat{J}_{1}^{\prime }(t_{1})\otimes t_{H}^{2}\frac{
e^{\phi ^{\prime }}}{p^{+}}\partial \delta (\sigma -\sigma ^{\prime })
    \notag \\
&&+a(t_{1})t_{P}^{2}\otimes \hat{J}_{1}(t_{2})\frac{e^{\phi }}{p^{+}}\partial
\delta (\sigma -\sigma ^{\prime })+a(t_{2})\hat{J}_{1}^{\prime }(t_{1})\otimes
t_{P}^{2}\frac{e^{\phi ^{\prime }}}{p^{+}}\partial \delta (\sigma -\sigma
^{\prime })\notag \\
&&-b(t_{1})t_{P}^{2}\otimes \hat{J}_{0}(t_{2})e^{-\phi }\partial \delta (\sigma
-\sigma ^{\prime })-b(t_{2})\hat{J}_{0}^{\prime }(t_{1})\otimes t_{P}^{2}e^{-\phi
^{\prime }}\partial \delta (\sigma -\sigma ^{\prime }) \\ \label{diracbracket}
&&+\Lambda_{1}(t_1,t_2)e^{-\phi}e^{\phi^{\prime}}\partial
\delta(\sigma-\sigma^{\prime})+\Lambda_{2}(t_1,t_2)e^{-\phi^{\prime}}e^{\phi}\partial
\delta(\sigma-\sigma^{\prime}) \notag \\
&&+\left( \mathit{S}^{5}\right), \notag
\end{eqnarray}
where $\Lambda_{1,2}(t_1,t_2)$ are constants depending on the spectral
parameters $(t_{1},t_{2})$, and $\left( \mathit{S}^{5}\right) $ represents terms depending only
on $e^{\phi(\sigma)}$ and coordinates $u^{M}$ parameterizing
$S^{5}$. The explicit form of these terms is presented in the Appendix
{\bf E}. This algebra has a closed form structure similar to~(\ref{central}), but
in this case the non-ultralocal terms are multiplied by a weight
factor $e^{\phi}$. In the Dirac
bracket (\ref{diracbracket}), the terms involving the
$AdS_{5}$ variables have a nice algebraic structure whereas the other
terms $\left( \mathit{S}^{5}\right) $ have a complicated form (see Appendix {\bf E}),
which, however, can be simplified in some subsectors.
We could not find a simpler expression for this part in general,
although it may be possible to simplify this term by choosing a
different parametrization or by making a gauge transformation (as, for
example, is done in~\cite{Arutyunov:2004yx} where one makes a gauge transformation to get rid of
non-local fields in the Lax pair).\footnote{We note here that the $\left(
\mathit{S}^{5}\right)$ part involves (see Appendix {\bf E}) the matrix $U$ defined in~(\ref{Umatrix}) which can also be
expressed in terms of components of the
current $J_{1}(\sigma)$, namely, in terms of $A^6$ and $B^6$ given in~(\ref{Ai}),
(\ref{Bj}) respectively, as a path-ordered integral. The dependence on the weight
factor $e^{\phi}$ arises due to the presence of such a factor in
$A^6$. As a result, all the non-ultralocal terms are multiplied by
weight factors of the form $e^{\phi(\sigma)}$.}

\section{Conclusion}
\label{sec:conclusion}

In this paper, we have studied the algebra of flat currents for the string
on $AdS_{5}\times S^{5}$ background in the light-cone gauge with the
$\kappa$ symmetry fixed. We show that the currents form a closed
algebra and present some explicit calculations. We point out that all the non-ultralocal
terms are multiplied by weight factors $e^{\phi(\sigma)}$.
From earlier results for sigma models coupled to gravity via dilaton field, this
suggests that such terms are unlikely to lead to ambiguities when
integrated. Further investigation of such questions including the 
Yangian algebra associated with the system is presently under way and will be reported later.

\section*{Appendix A: Properties of $psu(2,2|4)$}
\label{sec:appendix-A}
In this section we discuss some of the essential properties of the
superalgebra $psu(2,2|4)$ \cite{Berkovits:1999zq, Bershadsky:1999hk,
  Kac:1977qb, DeWitt:1992cy, Cornwell:1989bx}. Since we are
interested in a supersymmetric field theory, we assume that the algebra is
defined on a Grassmann space, $psu(2,2|4;\mathbb{C}B_{L})$. We represent an
element of this superalgebra by an even supermatrix of the form
\begin{equation}
G=\left(
\begin{array}{cc}
A & X \\
Y & B
\end{array}
\right) ,  \label{mat1}
\end{equation}
where $A$ and $B$ are matrices with Grassmann even functions while $X$ and $
Y $ are those with Grassmann odd functions, each representing a $4\times 4$
matrix. (An odd supermatrix, on the other hand, has the same form, with $A$
and $B$ consisting of Grassmann odd functions while $X$and $Y$ consisting of
Grassmann even functions.)

An element $G$ (see \ref{mat1}) of the superalgebra $psu (2,2|4; \mathbb{C}
B_{L})$ is given by a 8 $\times$ 8 matrix, satisfying
\begin{align}
& GK+KG^{\ddagger}=0,  \label{antihermite} \\
& \mbox{tr}A=\mbox{tr}B=0,  \label{mat3}
\end{align}
where $K=\left(
\begin{array}{cc}
\Sigma & 0 \\
0 & I_{4}
\end{array}
\right)$ and $\Sigma= \sigma_{3}\otimes I_{2}$ with $I_{2},I_{4}$
representing the identity matrix in $2$ and $4$ dimensions respectively. The
$\ddagger$ is defined by
\begin{equation}
G^{\ddagger}=G^{\mathrm{T} \sharp},  \label{mat4}
\end{equation}
where $\mathrm{T}$ denotes transposition and $\sharp$ is a generalization of
complex conjugation which acts on the functions $c$ of the matrices as
\begin{equation}
c^{\sharp}=\left\{
\begin{array}{cl}
c^{\ast} & \mbox{(for $c$ Grassmann even)} \\
-ic^{\ast} & \mbox{(for $c$ Grassmann odd)}
\end{array}
\right. .  \label{mat5}
\end{equation}
The condition (\ref{antihermite}) can be written explicitly as
\begin{equation}
\Sigma A^{\dagger}+A\Sigma=0,\quad B^{\dagger}+B=0,\quad X-i\Sigma
Y^{\dagger }=0.  \label{mat5a}
\end{equation}

The essential feature of the superalgebra $psu (2,2|4)$ is that it admits a $
\mathbb{Z}_{4}$ automorphism such that the condition $\mathbb{Z}_{4} (H) = H$
determines the maximal subgroup to be $SO(4,1)\times SO(5)$ which leads to
the definition of the coset for the sigma model. (This is the generalization
of the $\mathbb{Z}_{2}$ automorphism of bosonic sigma models to the
supersymmetric case.) The $\mathbb{Z}_{4}$ automorphism $\Omega$ takes an
element of $psu (2,2|4)$ to another, $G \rightarrow \Omega (G)$, such that
\begin{equation}
\Omega (G) =\left(
\begin{array}{cc}
JA^{\mathrm{T}}J & -JY^{\mathrm{T}}J \\
JX^{\mathrm{T}}J & JB^{\mathrm{T}}J
\end{array}
\right),  \label{mat6}
\end{equation}
where $J=\left(
\begin{array}{cc}
0 & -1 \\
1 & 0
\end{array}
\right)$. It follows now that $\Omega^{4} (G) = G$.

Since $\Omega^{4}=1$, the eigenvalues of $\Omega$ are $i^{p}$ with $p
=0,1,2,3$. Therefore, we can decompose the superalgebra as
\begin{equation}
G= \mathcal{H}_{0}\oplus \mathcal{H}_{1}\oplus \mathcal{H}_{2}\oplus
\mathcal{H}_{3},  \label{dec}
\end{equation}
where $\mathcal{H}_{p}$ denotes the eigenspace of $\Omega$ such that if $
H_{p}\in \mathcal{H}_{p}$, then
\begin{equation}
\Omega(H_{p})= i^{p}H_{p}.  \label{eig}
\end{equation}
We have already noted that $\Omega (\mathcal{H}_{0}) = \mathcal{H}_{0}$
determines $\mathcal{H}_{0} = SO(4,1)\times SO(5)$. $\mathcal{H}_{2}$
represents the remaining bosonic generators of the superalgebra while $
\mathcal{H}_{1},\mathcal{H}_{3}$ consist of the fermionic generators of the
algebra. (In a bosonic sigma model, $\mathcal{H}_{0},\mathcal{H}_{2}$ are
represented respectively as $Q,P$.) The automorphism also implies that
\begin{equation}
\lbrack H_{p},H_{q}]\in \mathcal{H}_{p+q\mbox{(mod 4)}}.  \label{mat8}
\end{equation}

The space $\mathcal{H}_{p}$ is spanned by the generators $(t_{p})_{A}$ of
the superalgebra so that we can explicitly write
\begin{eqnarray}
G & = & (H_{p})^{A}(t_{p})_{A}  \notag \\
& = & (H_{0})^{m}(t_{0})_{m}+(H_{1})^{\alpha_{1}}
(t_{1})_{\alpha_{1}}+(H_{2})^{A}(t_{2})_{A}+(H_{3})^{\alpha_{2}}(t_{3})_{
\alpha_{2}},  \label{mat9}
\end{eqnarray}
where $A=(m,\alpha_{1},a,\alpha_{2})$ take values over all the generators of
the superalgebra, $(H_{0})^{m}$ and $(H_{2})^{A}$ are Grassmann even
functions, while $(H_{1})^{\alpha1}$ and $(H_{3})^{\alpha2}$ are Grassmann
odd functions. The generators satisfy the graded algebra $psu(2,2|4)$,
\begin{equation}
\lbrack(t_{p})_{A},(t_{q})_{B}]=f_{AB}^{\quad C}(t_{p+q})_{C},  \label{mat10}
\end{equation}
where $p+q$ on the right hand side is to be understood modulo $4$.

The Killing form (or the bilinear form) $\langle H_{p},H_{q}\rangle $ is
also $\mathbb{Z}_{4}$ invariant so that
\begin{equation}
\langle \Omega \left( H_{p}\right) ,\Omega \left( H_{q}\right) \rangle
=\langle H_{p},H_{q}\rangle .
\end{equation}
This implies that
\begin{equation}
i^{(p+q)}\langle H_{p},H_{q}\rangle =\langle H_{p},H_{q}\rangle ,
\label{killing}
\end{equation}
which leads to
\begin{equation}
\langle H_{p},H_{q}\rangle =0\quad \mathrm{unless}\quad p+q=0\;(\mathrm{mod}
4).  \label{form}
\end{equation}
Since the supertrace of a supermatrix $M$ is defined as
\begin{equation}
\mbox{str}(M)=\left\{
\begin{array}{cl}
\mathrm{tr}A-\mathrm{tr}B & \mbox{(if $M$ is an even supermatrix)} \\
\mathrm{tr}A+\mathrm{tr}B & \mbox{(if $M$ is an odd supermatrix)}
\end{array}
\right. ,  \label{mat11}
\end{equation}
and the metric of the algebra is defined as $G_{AB}=\mathrm{str}
((t_{p})_{A}(t_{q})_{B})$, the above relation also implies that only the
components $G_{mn},G_{ab},G_{\alpha _{1}\alpha _{2}}=-G_{\alpha _{2}\alpha
_{1}}$ of the metric are non-zero. The structure constants possess the
graded anti-symmetry property
\begin{equation}
f_{AB}^{D}G_{DC}=-(-)^{|A||B|}f_{BA}^{D}G_{DC}=-(-)^{|B||C|}f_{AC}^{D}G_{DB},
\end{equation}
where $|A|$ denotes the Garssmann parity of $A$, namely, $|A|$ is 0 when $A$
is $m$ or $a$, while $|A|$ is 1 when $A$ is $\alpha _{1}$ or $\alpha _{2}$.

\section*{Appendix B: Properties of $so(3,1)\oplus su(4)$}
\label{sec:appendix-B}
We list here some relations between the generators and the Cartan
1-forms of the $psu(2,2|4)$
algebra written in $so(3,1)\oplus su(4)$ and $so(4,1)\oplus so(5)$ basis.

\begin{eqnarray}
P^{a} &=&\hat{P}^{a}+\hat{J}^{4a},\quad K^{a}=\frac{1}{2}\left( -\hat{P}
^{a}+\hat{J}^{4a}\right),\quad D=-\hat{P}^{4}, \notag\\
&& \notag \\
J_{j}^{i} &=&-\frac{i}{2}\left( \gamma ^{A^{\prime }}\right)
_{j}^{i}P^{A^{\prime }}+\frac{1}{4}\left( \gamma ^{A^{\prime }B^{\prime
}}\right) _{j}^{i}J^{A^{\prime }B^{\prime }}, \notag\\
&& \notag \\
\hat{L}^{a}&=&L_{P}^{a}-\frac{1}{2}L^{a}_{K},\quad
\hat{L}^{4a}=L^{a}_{P}+\frac{1}{2}L^{a}_{K},\quad
\hat{L}^{4}=-L_{D},\quad \hat{L}^{ab}=L^{ab} \notag \\
&& \notag \\
L_{j}^{i} &=&\frac{i}{2}\left( \gamma
^{A^{\prime}}\right)_{j}^{i}L^{A^{\prime }}
-\frac{1}{4}\left( \gamma ^{A^{\prime}B^{\prime
}}\right) _{j}^{i}L^{A^{\prime}B^{\prime}}, \notag\\
&& \notag\\
L^{A^{\prime}}&=&-\frac{i}{2}\left(\gamma^{A^{\prime}}\right)^{i}_{j}L^{j}_{i},\quad
L^{A^{\prime}B^{\prime}}=\frac{1}{2}\left(\gamma^{A^{\prime}B^{\prime}}\right)^{i}_{j}L^{j}_{i}\notag
\end{eqnarray}

\section*{Appendix C: $\{A^{a},B^{b}\}_{D}$}
\label{sec:appendix-C}
\bigskip We list here all the results for $\{A^{a},B^{b}\}_{D}$.

\bigskip

\begin{enumerate}
\item $\{A^{1},B^{1}\}_{D}= \{A^{2},B^{1}\}_{D} =...=\{A^{6},B^{1}\}_{D}=0$.

\item $\{A^{1},B^{2}\}_{D}=\frac{e^{\phi (\sigma )}}{p^{+}}A^{1}(\sigma
)B^{5}(\sigma )\delta (\sigma -\sigma ^{\prime })$.

\item $\{A^{2},B^{2}\}_{D}=-\frac{e^{\phi (\sigma )}}{p^{+}}B^{2}A^{5}\delta
(\sigma -\sigma ^{\prime })-\frac{e^{\phi (\sigma ^{\prime })}}{p^{+}}\left[
A^{2}(\sigma )+A^{2}(\sigma ^{\prime })\right] \partial _{\sigma ^{\prime
}}\delta (\sigma -\sigma ^{\prime })$.

\item $\{A^{3},B^{2}\}_{D}=-\frac{e^{\phi (\sigma ^{\prime })}}{p^{+}}
A^{3}(\sigma ^{\prime })\partial _{\sigma ^{\prime }}\delta (\sigma -\sigma
^{\prime })$.

\item $\{A^{4},B^{2}\}_{D}=-\frac{e^{\phi (\sigma ^{\prime })}}{p^{+}}
A^{4}(\sigma ^{\prime })\partial _{\sigma ^{\prime }}\delta (\sigma -\sigma
^{\prime })$.

\item $\{A^{5},B^{2}\}_{D}=\frac{e^{2\phi (\sigma
    )}}{p^{+}}B^{2}(\sigma )\delta
(\sigma -\sigma ^{\prime })-\frac{e^{\phi (\sigma ^{\prime })}}{p^{+}}
A^{5}(\sigma ^{\prime })\partial _{\sigma ^{\prime }}\delta (\sigma -\sigma
^{\prime })$.

\item $\{A^{6},B^{2}\}_{D}=-\frac{e^{\phi (\sigma ^{\prime })}}{p^{+}}
A^{6}(\sigma ^{\prime })\partial _{\sigma ^{\prime }}\delta (\sigma -\sigma
^{\prime })$.

\item $\{A^{1},B^{3}\}_{D}=0$.

\item $\{A^{2},B^{3}\}_{D}=-\frac{e^{\phi (\sigma )}}{p^{+}}B^{3}(\sigma
)A^{5}(\sigma )\delta (\sigma -\sigma ^{\prime })-\frac{e^{\phi (\sigma
^{\prime })}}{p^{+}}A^{3}(\sigma )\partial _{\sigma ^{\prime }}\delta
(\sigma -\sigma ^{\prime })$.

\item $\{A^{3},B^{3}\}_{D}=0$.

\item $\{A^{4},B^{3}\}_{D}=\frac{e^{\phi (\sigma )}e^{\phi (\sigma
    ^{\prime })}}{
p^{+}}\partial _{\sigma ^{\prime }}\delta (\sigma -\sigma ^{\prime })$.

\item $\{A^{5},B^{3}\}_{D}=\frac{e^{2\phi (\sigma
    )}}{p^{+}}B^{3}(\sigma )\delta (\sigma -\sigma ^{\prime })$.

\item $\{A^{6},B^{3}\}_{D}=0$.

\item $\{A^{1},B^{4}\}_{D}=0$.

\item $\{A^{2},B^{4}\}_{D}=-\frac{e^{\phi (\sigma )}}{p^{+}}B^{4}(\sigma
)A^{5}(\sigma )\delta (\sigma -\sigma ^{\prime })-\frac{e^{\phi (\sigma
^{\prime })}}{p^{+}}A^{4}(\sigma )\partial _{\sigma ^{\prime }}\delta
(\sigma -\sigma ^{\prime })$.

\item $\{A^{3},B^{4}\}_{D}=\frac{e^{\phi (\sigma )}e^{\phi (\sigma
    ^{\prime })}}{p^{+}}\partial _{\sigma ^{\prime }}\delta (\sigma
    -\sigma ^{\prime })$.

\item $\{A^{4},B^{4}\}_{D}=0$.

\item $\{A^{5},B^{4}\}_{D}=\frac{e^{2\phi (\sigma
    )}}{p^{+}}B^{4}(\sigma )\delta (\sigma -\sigma ^{\prime })$.

\item $\{A^{6},B^{4}\}_{D}=0$.

\item $\{A^{1},B^{5}\}_{D}=0$.

\item $\{A^{2},B^{5}\}_{D}=-\frac{e^{\phi (\sigma )}}{p^{+}}A^{5}(\sigma
)\partial _{\sigma ^{\prime }}\delta (\sigma -\sigma ^{\prime })$.

\item $\{A^{3},B^{5}\}_{D}=0$.

\item $\{A^{4},B^{5}\}_{D}=0$.

\item $\{A^{5},B^{5}\}_{D}=\frac{e^{2\phi (\sigma )}}{p^{+}}\partial _{\sigma
^{\prime }}\delta (\sigma -\sigma ^{\prime })$.

\item $\{A^{6},B^{5}\}_{D}=0$.

\item $\{A^{1},B^{6}\}_{D}=0$.

\item $\{A^{2},B^{6}\}_{D}=\frac{e^{\phi (\sigma )}}{p^{+}}\left(
B^{6}A^{6}-A^{6}B^{6}\right) \delta (\sigma -\sigma ^{\prime })+\frac{
e^{\phi (\sigma )}}{p^{+}}A^{6}(\sigma )\partial _{\sigma }\delta (\sigma
-\sigma ^{\prime })$.

\item $\{A^{3},B^{6}\}_{D}=\{A^{4},B^{6}\}_{D}=\{A^{5},B^{6}\}_{D}=0$.

\item $\{\left( A^{6}\right) _{j}^{i},\left( B^{6}\right)
  _{m}^{k}\}_{D}=\Omega
_{jm}^{ik}(\sigma )\partial _{\sigma ^{\prime }}\delta (\sigma -\sigma
^{^{\prime }})-\left( \Omega _{jn}^{ik}\left( B^{6}\right) _{m}^{n}-\left(
B^{6}\right) _{n}^{k}\Omega _{jm}^{in}\right) \delta (\sigma -\sigma
^{^{\prime }})$,

where
\begin{equation}
\Omega _{jm}^{ik}=-\frac{4e^{2\phi (\sigma )}}{p^{+}}\left[ \frac{\left(
\gamma ^{A}U^{-1}\right) _{j}^{i}}{Tr(U)}\frac{\left( \gamma
^{A}U^{-1}\right) _{m}^{k}}{Tr(U)}\right].  \label{omeg}
\end{equation}
\end{enumerate}
We note that we have used
\begin{align}
B^{6}& =\left( \partial _{\sigma }U\right) U^{-1},\quad A^{6}=\left( \partial
_{\tau }U\right) U^{-1},  \notag \\
&  \notag \\
U& \equiv \exp \left( \frac{i}{2}\gamma ^{A}y^{A}\right) =\cos \left(
\left\vert \frac{y}{2}\right\vert \right) +i\gamma ^{A}n^{A}\sin \left(
\left\vert \frac{y}{2}\right\vert \right),  \notag \\
&  \label{umatr2} \\
U^{-1}& =\exp \left( -\frac{i}{2}\gamma ^{A}y^{A}\right) =\cos \left(
\left\vert \frac{y}{2}\right\vert \right) -i\gamma ^{A}n^{A}\sin \left(
\left\vert \frac{y}{2}\right\vert \right),  \notag \\
&  \notag \\
n^{A}& =\frac{y^{A}}{\left\vert y\right\vert },\quad \left\vert y\right\vert =
\sqrt{y^{A}y^{A}},\quad \left( n^{A}\right) ^{2}=1,  \notag \\
&  \notag \\
\gamma ^{A}& =i\rho ^{A}\overline{\rho }^{6},\quad  A=1...5;i,j=1...4. \notag
\end{align}

\section*{Appendix D: $\{A^{a},A^{b}\}_{D}$}
\label{sec:appendix-D}
\bigskip We list here all the results for $\{A^{a},A^{b}\}_{D}$.

\begin{enumerate}
\item $\{A^{1},A^{1}\}_{D}=0$.

\item $\{A^{2},A^{1}\}_{D}=-\frac{e^{\phi (\sigma )}}{p^{+}}A^{1}(\sigma
)A^{5}(\sigma )\delta (\sigma -\sigma ^{\prime })$.

\item $\{A^{3},A^{1}\}_{D}=0$.

\item $\{A^{4},A^{1}\}_{D}=0$.

\item $\{A^{5},A^{1}\}_{D}=\frac{e^{2\phi (\sigma
    )}}{p^{+}}A^{1}(\sigma )\delta (\sigma -\sigma ^{\prime })$.

\item $\{A^{6},A^{1}\}_{D}=0$.

\item $\{A^{2},A^{2}\}_{D}=-\frac{e^{5\phi (\sigma )}}{\left(
  p^{+}\right) ^{3}}
B^{2}(\sigma )\partial _{\sigma ^{\prime }}\delta (\sigma -\sigma ^{\prime
})-\frac{e^{5\phi (\sigma ^{\prime })}}{\left( p^{+}\right) ^{3}}
B^{2}(\sigma ^{\prime })\partial _{\sigma ^{\prime }}\delta (\sigma -\sigma
^{\prime })$.

\item $\{A^{3},A^{2}\}_{D}=-\frac{e^{5\phi (\sigma ^{\prime })}}{\left(
p^{+}\right) ^{3}}B^{3}(\sigma ^{\prime })\partial _{\sigma ^{\prime
}}\delta (\sigma -\sigma ^{\prime })+\frac{e^{\phi (\sigma )}}{\left(
p^{+}\right) }A^{3}A^{5}\delta (\sigma -\sigma ^{\prime })$

$-\frac{e^{5\phi (\sigma )}}{\left( p^{+}\right) ^{3}}B^{3}B^{5}\delta
(\sigma -\sigma ^{\prime })$.

\item $\{A^{4},A^{2}\}_{D}=-\frac{e^{5\phi (\sigma ^{\prime })}}{\left(
p^{+}\right) ^{3}}B^{4}(\sigma ^{\prime })\partial _{\sigma ^{\prime
}}\delta (\sigma -\sigma ^{\prime })+\frac{e^{\phi (\sigma )}}{\left(
p^{+}\right) }A^{4}A^{5}\delta (\sigma -\sigma ^{\prime })$

$-\frac{e^{5\phi (\sigma )}}{\left( p^{+}\right) ^{3}}B^{4}B^{5}\delta
(\sigma -\sigma ^{\prime })$.

\item $\{A^{5},A^{2}\}_{D}=3\frac{e^{2\phi (\sigma
    )}}{p^{+}}A^{2}\delta (\sigma
-\sigma ^{\prime })+2\frac{e^{\phi (\sigma )}}{p^{+}}A^{3}A^{4}\delta
(\sigma -\sigma ^{\prime })$

$-2\frac{e^{5\phi (\sigma )}}{\left( p^{+}\right) ^{3}}B^{3}B^{4}\delta
(\sigma -\sigma ^{\prime })+2\frac{e^{\phi (\sigma )}}{p^{+}}\left(
A^{5}\right) ^{2}\delta (\sigma -\sigma ^{\prime })-\frac{e^{5\phi (\sigma
^{\prime })}}{\left( p^{+}\right) ^{3}}B^{5}(\sigma ^{\prime })\partial
_{\sigma ^{\prime }}\delta (\sigma -\sigma ^{\prime })$

$-2\frac{e^{5\phi (\sigma )}}{\left( p^{+}\right) ^{3}}\left( B^{5}\right)
^{2}\delta (\sigma -\sigma ^{\prime })$.

\item $\{A^{6},A^{2}\}_{D}=-\frac{e^{6\phi (\sigma ^{\prime })}}{\left(
p^{+}\right) ^{3}}B^{6}(\sigma ^{\prime })\partial _{\sigma ^{\prime
}}\delta (\sigma -\sigma ^{\prime })+2\frac{e^{2\phi (\sigma )}}{\left(
p^{+}\right) }\left( A^{5}A^{6}-B^{5}B^{6}\right) \delta (\sigma -\sigma
^{\prime })$

$+\frac{e^{2\phi (\sigma )}}{2\left( p^{+}\right) }\left[ \frac{
A^{6}U^{-1}A^{6}U^{-1}+A^{6}U^{-2}A^{6}+\left( A^{6}\right)
^{2}U^{2}+UA^{6}U^{-1}A^{6}U^{2}}{(1+u^{6})}\right] \delta (\sigma -\sigma
^{\prime })$

$-\frac{e^{2\phi (\sigma )}}{2\left( p^{+}\right) }\left[ \frac{
B^{6}U^{-1}B^{6}U^{-1}+B^{6}U^{-2}B^{6}+\left( B^{6}\right)
^{2}U^{2}+UB^{6}U^{-1}B^{6}U^{2}}{(1+u^{6})}\right] \delta (\sigma -\sigma
^{\prime })$.

\item $\{A^{3},A^{3}\}_{D}=0$.

\item $\{A^{4},A^{3}\}_{D}=0$.

\item $\{A^{5},A^{3}\}_{D}=\frac{e^{2\phi (\sigma
    )}}{p^{+}}A^{3}(\sigma )\delta (\sigma -\sigma ^{\prime })$.

\item $\{A^{6},A^{3}\}_{D}=0$.

\item $\{A^{4},A^{4}\}_{D}=0$.

\item $\{A^{5},A^{4}\}_{D}=\frac{e^{2\phi (\sigma
    )}}{p^{+}}A^{4}(\sigma )\delta (\sigma -\sigma ^{\prime })$.

\item $\{A^{6},A^{4}\}_{D}=0$.

\item $\{A^{5},A^{5}\}_{D}=0$.

\item $\{A^{6},A^{5}\}_{D}=-2\frac{e^{2\phi (\sigma )}}{p^{+}}A^{6}(\sigma
)\delta (\sigma -\sigma ^{\prime })$.

\item  $\{\left( A^{6}\right) _{j}^{i},\left( A^{6}\right)
  _{m}^{k}\}_{D} =\frac{
e^{2\phi (\sigma )}}{2p^{+}}\left[
-\frac{\left[ \left( A^{6}U+UA^{6}\right) U\gamma ^{A}\right] _{j}^{i}\left[
\gamma ^{A}U^{-1}\right] _{m}^{k}}{\sqrt{2}\left( 1+u^{6}\right) ^{3/2}}+
\frac{\left[ \left( A^{6}U+UA^{6}\right) U\gamma ^{A}\right] _{m}^{k}\left[
\gamma ^{A}U^{-1}\right] _{j}^{i}}{\sqrt{2}\left( 1+u^{6}\right) ^{3/2}}
\right. \\
\left. \qquad \qquad \qquad \qquad +\frac{\left( A^{6}\right) _{j}^{i}\left( U^{-1}\right) _{m}^{k}}{\sqrt{2}
\sqrt{\left( 1+u^{6}\right) }}-\frac{\left( A^{6}\right) _{m}^{k}\left(
U^{-1}\right) _{j}^{i}}{\sqrt{2}\sqrt{\left( 1+u^{6}\right) }}\right]
  \delta (\sigma -\sigma ^{\prime }).$
\end{enumerate}

In relations 11 and 21 one should replace $(1+u^{6})$ using one of the
following expressions:
\begin{equation}
\sqrt{(1+u^{6})}=\frac{\sqrt{2}}{4}Tr(U),  \label{mess2}
\end{equation}
or
\begin{equation}
1+u^{6}=\frac{1}{2}\left( U+U^{-1}\right) ^{2}.  \label{mess3}
\end{equation}

\section*{Appendix E: Expression for $(S^{5})$}
\label{sec:appendix-E}
\bigskip
\begin{eqnarray*}
\Lambda_{1}(t_{1},t_{2}) &&=b(t_{1})(t_{P}^{1}\otimes t_{H}^{2}+t_{P}^{2}\otimes
t_{H}^{1}+t_{P}^{3}\otimes t_{H}^{4}+t_{P}^{4}\otimes t_{H}^{3})\notag \\
&&+b(t_{1})a(t_{2})(t_{P}^{1}\otimes t_{P}^{2}+t_{P}^{2}\otimes
t_{P}^{1}+t_{P}^{3}\otimes t_{P}^{4}+t_{P}^{4}\otimes
t_{P}^{3}+t_{P}^{5}\otimes t_{P}^{5})\notag \\
\end{eqnarray*}

\begin{eqnarray*}
\Lambda_{2}(t_{1},t_{2}) &&=b(t_{2})(t_{H}^{1}\otimes t_{P}^{2}+t_{H}^{2}\otimes
t_{P}^{1}+t_{H}^{3}\otimes t_{P}^{4}+t_{H}^{4}\otimes t_{P}^{3})\notag\\
&&+b(t_{2})a(t_{1})(t_{P}^{1}\otimes t_{P}^{2}+t_{P}^{2}\otimes
t_{P}^{1}+t_{P}^{3}\otimes t_{P}^{4}+t_{P}^{4}\otimes
t_{P}^{3}+t_{P}^{5}\otimes t_{P}^{5})\notag\\
\end{eqnarray*}

\begin{eqnarray*}
&&\left( \mathit{S}^{5}\right) =b(t_{1})t_{p}^{5}\otimes \left(
B^{6}t_{H}^{6}+a(t_{2})B^{6}t_{P}^{6}-b(t_{2})
\bar{A}^{6}t_{P}^{6}\right) \delta (\sigma -\sigma ^{\prime })
\\
&&-b(t_{2})\left(
B^{6}t_{H}^{6}+a(t_{1})B^{6}t_{P}^{6}-b(t_{1})
\bar{A}^{6}t_{P}^{6}\right) \otimes t_{P}^{5}\delta (\sigma
-\sigma ^{\prime }) \\
&&+b(t_{1})t_{P}^{2}\otimes (\bar{A^{6}}_{m}^{k}(B^{6})_{n}^{j}\rho
^{mn}\rho _{ki}(t_{H}^{6})_{j}^{i})\frac{e^{\phi }}{p^{+}}\delta (\sigma
-\sigma ^{\prime }) \\
&&-b(t_{2})(\bar{A^{6}}_{m}^{k}(B^{6})_{n}^{j}\rho ^{mn}\rho
_{ki}(t_{H}^{6})_{j}^{i})\otimes t_{P}^{2}\frac{e^{\phi }}{p^{+}}\delta
(\sigma -\sigma ^{\prime }) \\
&&-\frac{1}{2}b(t_{1})a(t_{2})t_{P}^{2}\otimes [\bar{A^{6}}
_{m}^{k}(B^{6})_{n}^{j}\rho ^{mn}\rho _{ki}-(B^{6})_{m}^{k}\bar{A^{6}}
_{n}^{j}\rho ^{mn}\rho _{ki}+\bar{A^{6}}_{i}^{m}(B^{6})_{k}^{n}\rho
_{mn}\rho ^{kj} \\
&&-(B^{6})_{i}^{m}\bar{A^{6}}_{k}^{n}\rho _{mn}\rho ^{kj}](t_{P}^{6})_{j}^{i}
\frac{e^{\phi }}{p^{+}}\delta (\sigma -\sigma ^{\prime }) \\
&&\qquad \qquad \qquad \qquad \\
&&+\frac{1}{2}b(t_{2})a(t_{1})[\bar{A^{6}}_{m}^{k}(B^{6})_{n}^{j}\rho
^{mn}\rho _{ki}-(B^{6})_{m}^{k}\bar{A^{6}}_{n}^{j}\rho ^{mn}\rho _{ki}+\bar{
A^{6}}_{i}^{m}(B^{6})_{k}^{n}\rho _{mn}\rho ^{kj} \\
&&-(B^{6})_{i}^{m}\bar{A^{6}}_{k}^{n}\rho _{mn}\rho
^{kj}](t_{P}^{6})_{j}^{i}\otimes t_{P}^{2}\frac{e^{\phi }}{p^{+}}\delta
(\sigma -\sigma ^{\prime }) \\
&&\qquad \qquad \qquad \qquad \\
&&+b(t_{1})b(t_{2})\left[ \frac{1}{1+u^{6}}(\bar{A}^{6}U^{-1}\bar{A}
^{6}U^{-1}+\bar{A}^{6}U^{-2}\bar{A}^{6}+(\bar{A}^{6})^{2}U^{2}+U\bar{A}
^{6}U^{-1}\bar{A}^{6}U^{2}\right. \\
&&\quad
  \left.-B^{6}U^{-1}B^{6}U^{-1}-B^{6}U^{-2}B^{6}-(B^{6})^{2}U^{2}
-UB^{6}U^{-1}B^{6}U^{2})_{j}^{i}(t_{P}^{6})_{i}^{j}\otimes,t_{P}^{2}
\right] \frac{e^{\phi }}{2p^{+}}\delta (\sigma -\sigma ^{\prime }) \\
&&+b(t_{1})\{(\bar{A^{6}})_{j}^{i},(B^{\prime
}{}^{6})_{l}^{k}\}(t_{P}^{6})_{i}^{j}\otimes
(t_{H}^{6})_{k}^{l}+b(t_{2})\{(B^{6})_{l}^{k},(\bar{A^{\prime }{}^{6}}
)_{j}^{i}\}(t_{H}^{6})_{k}^{l}\otimes (t_{P}^{6})_{i}^{j} \\
&&+b(t_{1})a(t_{2})\{(\bar{A^{6}})_{j}^{i},(B^{\prime
}{}^{6})_{l}^{k}\}(t_{P}^{6})_{i}^{j}\otimes
(t_{P}^{6})_{k}^{l}+b(t_{2})a(t_{1})\{(B^{6})_{l}^{k},(\bar{A^{\prime }{}^{6}
})_{j}^{i}\}(t_{P}^{6})_{k}^{l}\otimes (t_{P}^{6})_{i}^{j} \\
&&+b(t_{1})b(t_{2})\{(\bar{A^{6}})_{j}^{i},(\bar{A}^{\prime
}{}^{6})_{l}^{k}\}(t_{P}^{6})_{i}^{j}\otimes (t_{P}^{6})_{k}^{l}.
\end{eqnarray*}

\vskip.7cm \noindent \textbf{Acknowledgement:}

One of us (A.M.) would like to thank Abhishek Agarwal for many helpful
discussions. M.S. would also like to thank A. Agarwal, T. Asakawa, D.
Korotkin, K. Ohta, Y. Okawa, K. Yoshida  for useful discussions.
This work is supported in part by US DOE Grant No. DE-FG 02-91ER40685.

\bibliographystyle{JHEP3}
\bibliography{lightcone_final}

\end{document}